# Image Sterilization to Prevent LSB-based Steganographic Transmission

Goutam Paul[1], *Member*, *IEEE*, and Imon Mukherjee[2]

***Abstract*—Sterilization is a very popular word used in biomedical testing (like removal of all microorganisms on surface of an article or in fluid using appropriate chemical products). Motivated by this biological analogy, we, for the first time, introduce the concept of sterilization of an image, i.e., removing any steganographic information embedded in the image. Experimental results show that our technique succeeded in sterilizing around 76% to 91% of stego pixels in an image on average, where data is embedded using LSB-based steganography.**

## I. Introduction and Motivation

*Steganography* [1] is the popular word that takes birth from a Greek word "Steganos", referring to "hiding information" in innocuous media like images. The steganographic technique hides messages inside other media or messages so that the hidden message will not be detectable by enemy or any other person in normal sense.

The Germans have successfully utilized this technology in the Second World War [2]. Academic research in steganography has grown tremendously in last few decades and many steganographic algorithms exist in the literature. An image or media before embedding any secret information is called *cover* and after inserting information is called *stego*.

*Steganalysis* [3][6] is the art and science to defeat steganography. Steganalysis is done neither having the prior knowledge of secret key used for embedding the information into the cover media nor knowing the steganographic algorithm used. That is why determining whether the secret message exists in the media is difficult and challengeable task.

Steganography can be applied to a variety of multimedia contents like images, audio, video, text etc. We focus on image steganography in this paper. Our objective in this work is to develop an algorithm to revert as many stego-pixels of an image as possible to their original cover form. We call this *image sterilization*.

Image sterilization may have important application in defense and security domain. For example, suppose that a spy wants to inform his team about the venue of making a bomb blast in some target place using some image based steganographic technique. During the time of transmission, if sterilization of the stego information is performed by the security personals, then the attackers would be completely unaware about the venue and their plan may be jeopardized.

## II. Steganographic Techniques in Spatial Domain

Before going into the details of our sterilization technique, we briefly describe here the major categories of steganographic algorithms in the spatial domain.

The steganographic algorithms operating in the spatial domain as the method for selecting the pixels can be classified into three categories: non-filtering algorithms, randomized algorithms and filtering algorithms [9].

### *A. Non-filtering Algorithm*

The non-filtering steganographic algorithm [9] is the most popular and the most vulnerable steganographic technique based on LSB. The embedding process is done as a sequential substitution of each LSB of the image pixel for each bit of the message. Hence a large amount of information can be stored into such cover media.

The only requirement is sequential LSB reading, starting from the first image pixel in order to extract the secret message from the cover media (viz. image). As the message is embedded at the initial pixels of the image, leaving the remaining pixels unchanged, this technique is nothing but an unbalanced distribution of the changed pixels.

[1] Dept. of Computer Science & Engg., Jadavpur University, Kolkata, India, goutam.paul@ieee.org

[2] Dept. of Computer Science & Engg., Inst. of Technolgy and Marine Engg., S. 24 Parganas, India, mukherjee.imon@gmail.com

*B. Randomized Algorithm*

This technique solves the limitation of the previous technique. Each of the sender and the receiver has a password denominated stego-key which is generated through a pseudo-random number generator [9]. This creates an index sequence to access the image pixel. The message bit is embedded in the pixel of the cover media following the index sequence produced by the pseudo-random number generator

The two main features of this technique are: a) use of password to have access to the message and b) well-spread message bits over the image which is difficult to detect compare to the previous one.

*C. Filtering Algorithm*

The filtering algorithm [9] filters the cover image by using a default filter and hides information in those areas that get a better rate. The filter is used to the most significant bits of every pixel, leaving the less significant bits to hide information. The filter gives the guarantee of a greater difficulty of detecting the presence of hidden messages. The retrieval of information is ensured because the bits used for filtering are not changed.

Each of the aforesaid three categories of steganographic techniques in spatial domain is susceptible to image sterilization described in subsequent sections.

## III. Image Sterilization

The 24-bit color image consists of a number of pixels and each pixel contains three intensity levels (of 8 bits each), one for each of red, green and blue color components.

One of the most popular steganography techniques is Least Significant Bit (LSB) insertion [4]. Typically, there are thousands of pixels in an image. So if we change the LSB of some pixels, the resulting picture will probably be alike to the original image.

---

128 64 32 16 8 4 2 1

---

⇩ MSB ⇩ LSB

The LSB flipping function [5] for a stego image is defined by $F_1 = 0 \leftrightarrow 1,\ 2 \leftrightarrow 3,\ 4 \leftrightarrow 5, \ldots$, etc. The groups are formed based on this flipping function. The intensity values $2j$ and $2j+1$, for $0 \le j \le 127$, belong to the same group. So the maximum possible number of groups for each component of an image is 128. Suppose that an image contains $N$ pixels with $c$ groups. Let $n_i$ be the number of pixels in the $i^{th}$ group, $1 \le i \le c$. Thus, $N = \sum n_i$. The set of pixels (based on their intensity values) for the $i^{th}$ group is represented by $G_i = \{x_{i,k} : 1 \le k \le n_i\}$, such that $x_{i,j} - x_{i,m} \in \{-1,0,+1\}$, $1 \le j \ne m \le n_i$.

| | | | | | | | | |
|---|---|---|---|---|---|---|---|---|
| 48 | 48 | 49 | 48 | 48 | 77 | 76 | 76 | 77 |
| 48 | 49 | 49 | 49 | 48 | 48 | 48 | 77 | 77 |
| 48 | 48 | 77 | 77 | 76 | 77 | 76 | 76 | 77 |
| 48 | 76 | 76 | 76 | 77 | 77 | 77 | 48 | 49 |
| 48 | 77 | 77 | 76 | 77 | 77 | 76 | 48 | 49 |
| 48 | 48 | 49 | 77 | 77 | 77 | 77 | 49 | 49 |
| 49 | 49 | 48 | 77 | 77 | 77 | 77 | 49 | 48 |
| 49 | 49 | 49 | 49 | 77 | 77 | 77 | 77 | 49 |

Figure 1. Sample intensity matrix for a single component of an image and the corresponding two groups

In Fig. 1, we show some sample intensity values of one component (either blue, red or green) of an arbitrary image. There are two groups – one shown in red boundary, another in blue. Suppose that the intensity of a pixel in the original image is 48. After stego insertion, the value may either remain as 48 (if 0 is inserted) or be changed to 49 (if 1 is inserted, as shown in Fig. 2 below). So we have considered 48 and 49 in the same group.

48 = 0 0 1 1 0 0 0 [0]

⇩

0 0 1 1 0 0 0 [1] = 49

Figure 2. LSB embedding in a pixel

In Algorithm 1, we present our procedure for image sterilization, called *SterilizeStegoImage*.

ALGORITHM 1: SterilizeStegoImage

**Input**: A stego-image.
**Output**: The sterilized image.

**Step-1**: Read the intensity values from the stego image.
**Step-2**: For each color component
  **Step-2.1**: Form the groups based on the LSB flipping function.
  **Step-2.1**: For each group,
    **Step 2.1.1**: Count the *odd* and *even* pixels with intensity values of the form $2j+1$ and $2j$ respectively; Let $n_o$ and $n_e$ be the respective counts.
    **Step 2.1.2**: If $n_e > n_o$, then replace all $2j+1$ intensity values by $2j$.
    **Step 2.1.3** Else replace all $2j$ intensity values by $2j+1$.
**Step-3**: Output the transformed image.

The basic idea behind our image sterilization technique is to replace an intensity value X with some other value Y such that one cannot extract the hidden message from the cover media. This technique can be applied to both 24-bit color images as well as 8-bit grayscale images. Each group contains at most two intensity values, of the form 2*j* (we call them *even* pixels) and 2*j*+1 (we call them *odd* pixels). Let $n_o$ and $n_e$ be the number of *even* and *odd* pixels in a group. If $n_e > n_o$, we replace all 2*j*+1 intensity values by 2*j*, otherwise we do the opposite replacement. In other words, we force all pixels in a group to be either *odd* or *even* depending on the majority of the pixels being *odd* or *even*.

## IV. ACCURACY MEASUREMENT

To estimate the accuracy of our technique, we need to take as inputs some sample stego images for which we know which pixel values are actually changed due to the LSB embedding. Let S be the number of stego pixels and $S^{/}$ out of those S pixels actually differ in intensity values when compared with the corresponding cover image. Now, suppose $S^{//}$ out of those $S^{/}$ pixels are recovered due to the sterilization process. We calculate the accuracy of sterilization for this image as $S^{//}/S^{/}$.

We have used a database of fifty 24-bit color images in BMP format on natural scenario and fifty gray scale images (downloaded from [7]). We have also prepared 100 different text files containing the story of Sherlock home's (downloaded from [8]). Each pixel of a 24-bit color image contains three components, viz. red, green and blue. So using LSB embedding, at the most three bits of data can be embedded in a pixel. If the dimension of an image is $m \times n$ then maximum number of data bits possible to be inserted in the 24 bit color image can be $m \times n \times 3$. Since a character is of eight bit, the maximum number of characters (including space) of the text would be $m \times n \times 3/8$. Thus, each of the 50 text files consists of at most $m \times n \times 3/8$ characters depending upon the values of the dimension *m* and *n*. Similarly for gray scale image each text file should have $m \times n/8$ characters. We have used MATLAB 7.7.0 as a software tool for implementation.

Below we give an example of how the text extracted from a sterilized image may differ from the original text that was embedded and is expected to be extracted from the unsterilized stego counterpart.

| |
|---|
| I was the means of introducing to his notice that of Mr Hatherleys thumb and that of Colonel Warburtons madness. |

Embeded Messege(using [16]) before Sterilization

⇩

| |
|---|
| J%ybtsgf%ofbmt$nmssnevcjnhp%gjt$mnshb dsgbt$nh$Ns%Gbsgfqmfyt$sgvnb%bme$sh? u$nf$Bnmpmfm%X?savasnmt$nbdmftr/ |

Embeded Messege(using proposed algorithm) after Sterilization

Fig. 3 and Fig. 4 show a stego and the corresponding sterilized gray scale image and Fig. 5 and Fig. 6 show a stego and the corresponding sterilized 24-bit color image respectively.

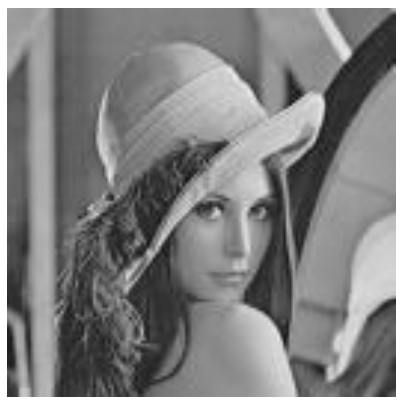

Figure 3. A sample gray scale stego image ("lena.bmp")

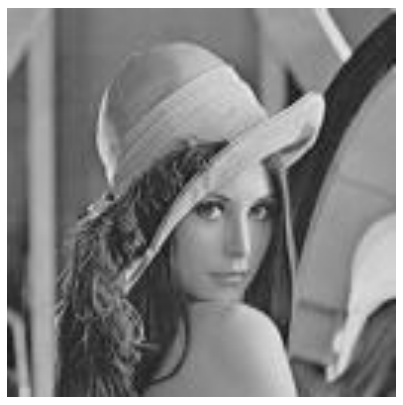

Figure 4. The sterilized version of Fig. 3 ("lena_steri.bmp")

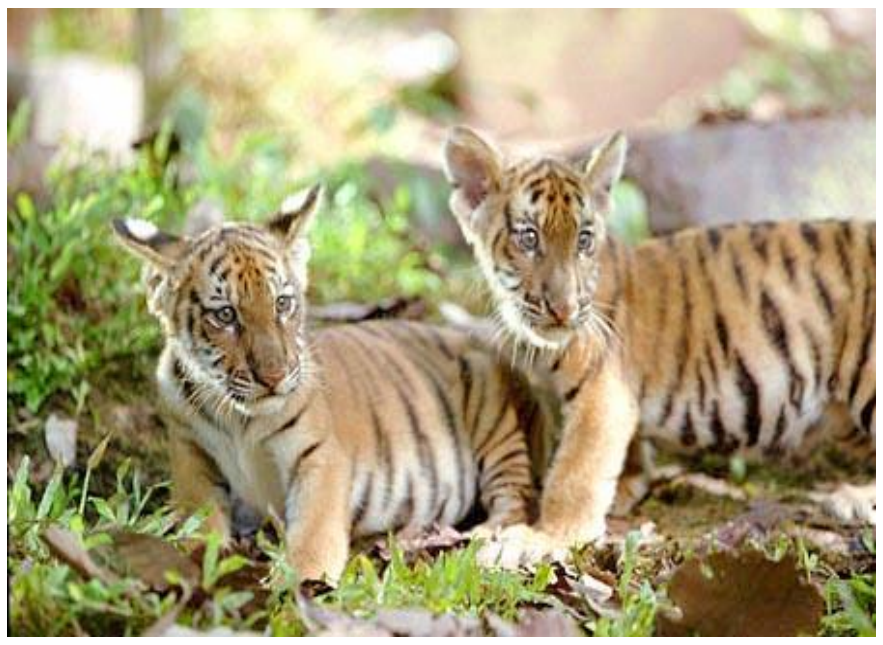

Figure 5. A sample 24-bit color stego image ("cubs24.bmp")

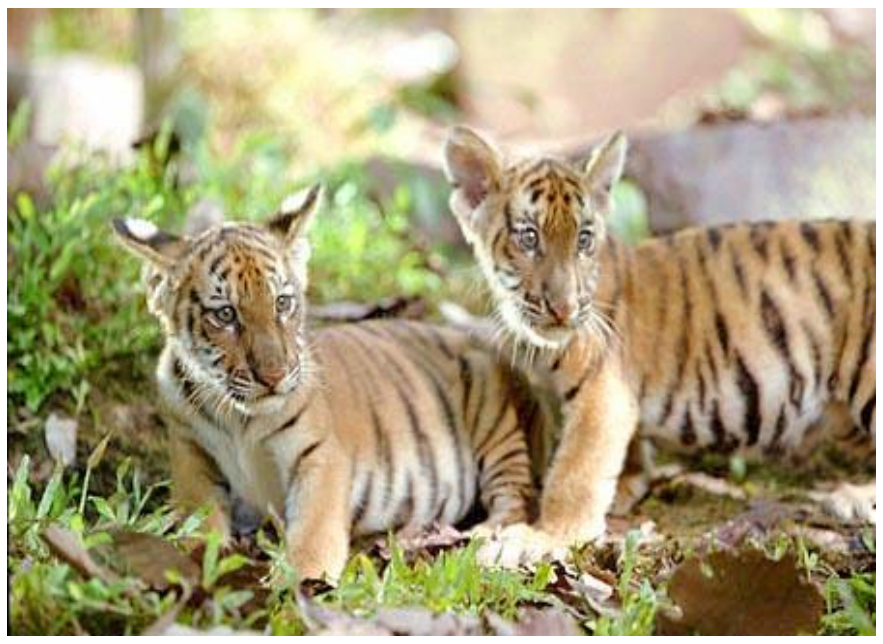

Figure 6. The sterilized version of Figure 5

We observe that after sterilization, the actual message is scrambled enough so that it cannot be reliably recovered.

Table I shows the ability to sterilize the stego image using the algorithm described in the previous section for three different embedding techniques. The first algorithm is the naïve LSB based sequential embedding technique; we call it algorithm A. Another technique is taken from [15], which we refer as algorithm B. The third method, denoted by algorithm

C, uses random pixel selection and segmentation mechanisms [16].

TABLE I: ACCURACY (MINIMUM, MAXIMUM, AVERAGE AND STANDARD DEVIATION) OF STERILIZATION OVER FIFTY GRAY SCALE AND FIFTY 24-BIT COLOR IMAGES FOR THREE DIFFERENT ALGORITHMS A, B, C.

| Image Type / Accuracy | | Gray Scale Image | 24-Bit color Image | | |
|---|---|---|---|---|---|
| | | | R | G | B |
| Minimum % | A | 72.5 | 68.01 | 68.9 | 69.75 |
| | B | 79.27 | 74.41 | 76.2 | 75.57 |
| | C | N.A | 84.29 | 84.36 | 85.89 |
| Avg % | A | **78.09** | **77.16** | **76.64** | **78.34** |
| | B | **80.57** | **79.56** | **78.78** | **80.49** |
| | C | **N.A** | **91.03** | **89.88** | **91.37** |
| Maximum % | A | 87.74 | 90.85 | 91.01 | 90.02 |
| | B | 83.60 | 88.6 | 82.72 | 91.44 |
| | C | N.A | 96.12 | 93.38 | 90.76 |
| Standard Deviation | A | 0.0351 | 0.0622 | 0.0769 | 0.0729 |
| | B | 0.0222 | 0.0483 | 0.0226 | 0.0562 |
| | C | N.A | 0.0392 | 0.0336 | 0.0431 |

Since a color image has three components, we have separately measured the performance for each component. The average accuracy (computed over 50 images) for the red, green and blue components for algorithms A, B, C are marked bold. The low standard deviations indicate that our estimate is robust.

## V. OTHER PERFORMANCE PARAMETERS

### A. *Mean Squared ERROR (MSE) and Peak Signal to Noise Ratio(PSNR)*

The imperceptibility of hidden information in an image is measured by stego image quality in terms of Mean Square Error (MSE) and Peak-Signal-to-Noise Ratio (PSNR) in dB [10] [12].

Consider a discrete image $A(m, n)$ for $m=1, 2, \ldots, M$ and $n=1, 2, \ldots, N$, which is considered as a reference image.

Consider a second image $B(m, n)$, of the same spatial dimension as $A(m, n)$, that is to be compared to the reference image.

MSE is given as

$$\text{MSE}= \frac{\sum_{M,N}\left(A(m,n)-B(m,n)\right)^2}{MN},$$

where, $M$ and $N$ are the number of rows and columns in the input images and PSNR is given by

$$\text{PSNR}= 10 \log_{10} \frac{255^2}{\text{MSE}}.$$

The MSE represents the cumulative squared error between the two images. The mean square error measure is very popular because it can correlate reasonably with subjective visual tests and it is mathematically tractable.

Lower MSE and higher PSNR imply that the difference between the original image and test image is small, i.e., it is usually not possible to distinguish whether the image is a stego one or a sterilized one. In our experiments, we have obtained very low MSE (0.0050 for gray scale image and 0.0052 for color image). Similarly the PSNR of cover and sterilized images are very high (71.18 dB for gray scale image and 75.05 dB for color image). These results indicate that our technique is successful in hiding the fact that the image has been sterilized.

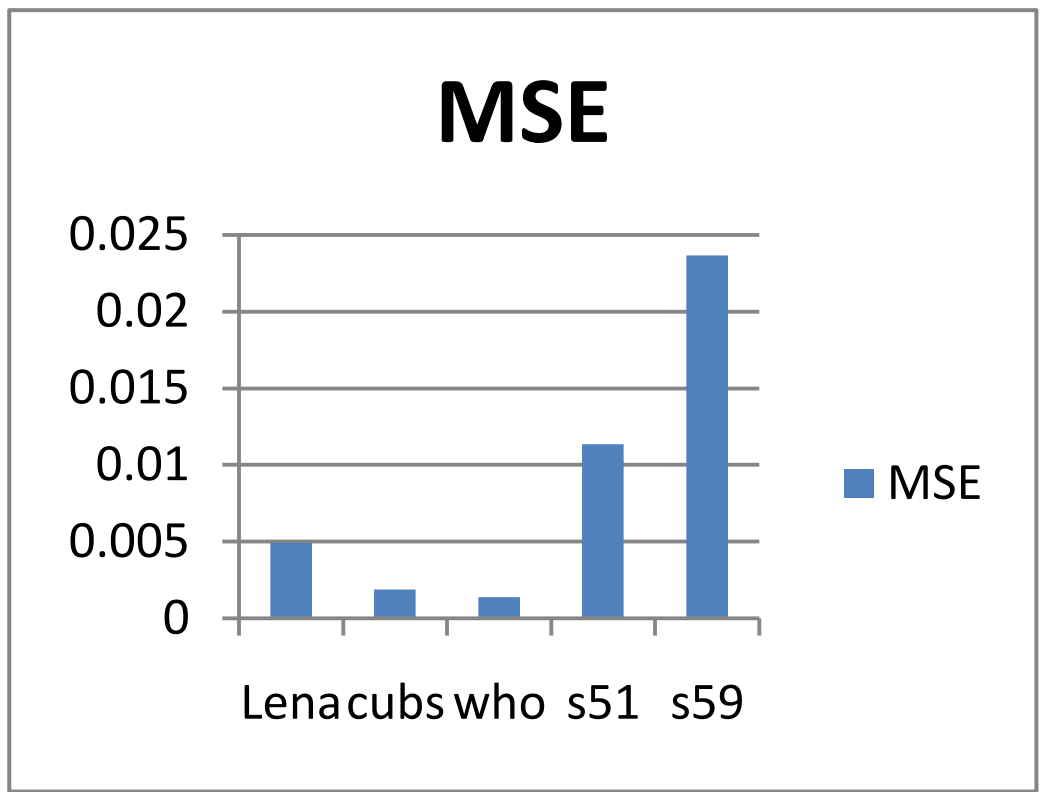

Figure 7. Mean Squared Error of some randomly chosen sample images out of 50 images ( Lena.bmp, Cubs.bmp, WHO.bmp, s51.bmp and s59.bmp)

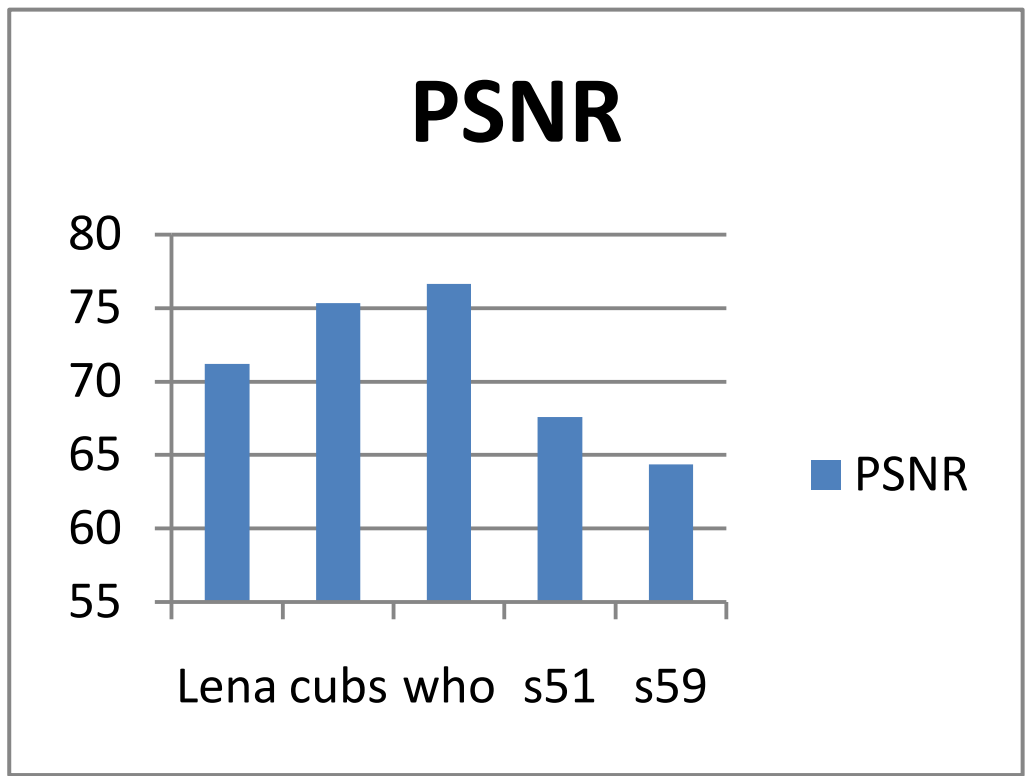

Figure 8. Peak Signal to Noise Ratio of some randomly chosen sample images out of 50 images ( Lena.bmp, Cubs.bmp, WHO.bmp, s51.bmp and s59.bmp)

### B. *Histogram Analysis:*

The main purpose of histogram analysis [11] in our context is to detect significant changes in frequency of appearance of the each color component in an image by comparing the cover image with the steganographic image and sterilized image.

Fig. 9, Fig. 10 and Fig. 11 show the histogram of the Lena image in three stages: before stego insertion, after stego insertion and after sterilization.

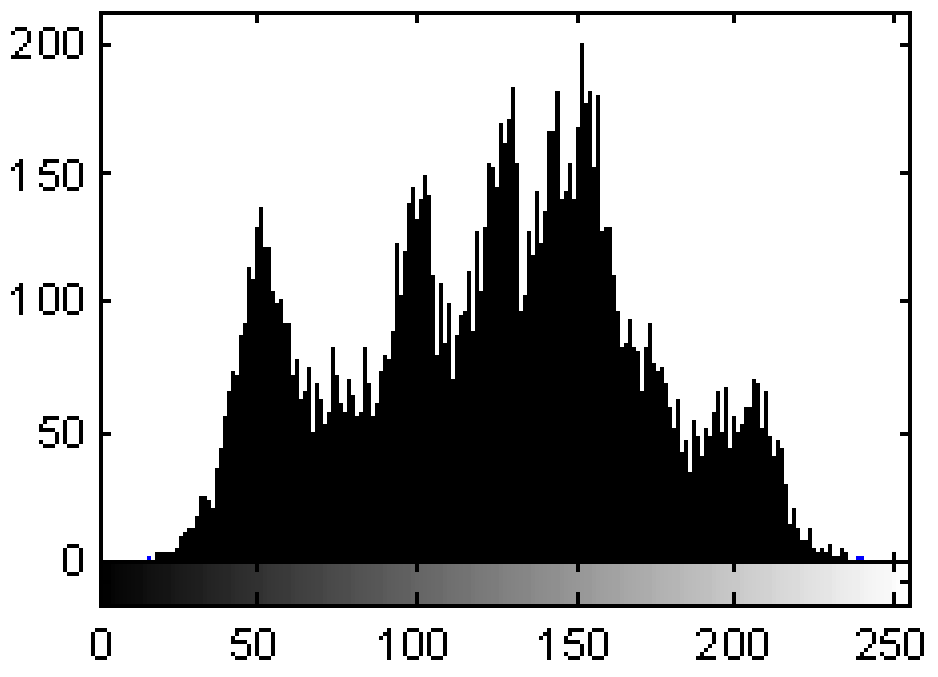

Figure 9. Histogram of original Lena image

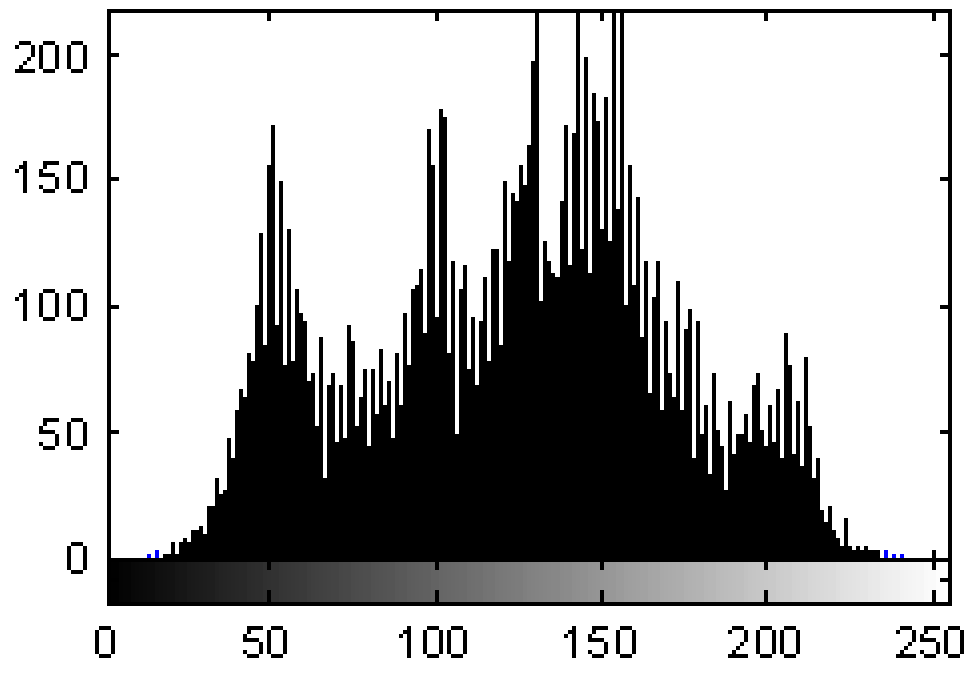

Figure 10. Histogram of Lena image

(After stego insertion)

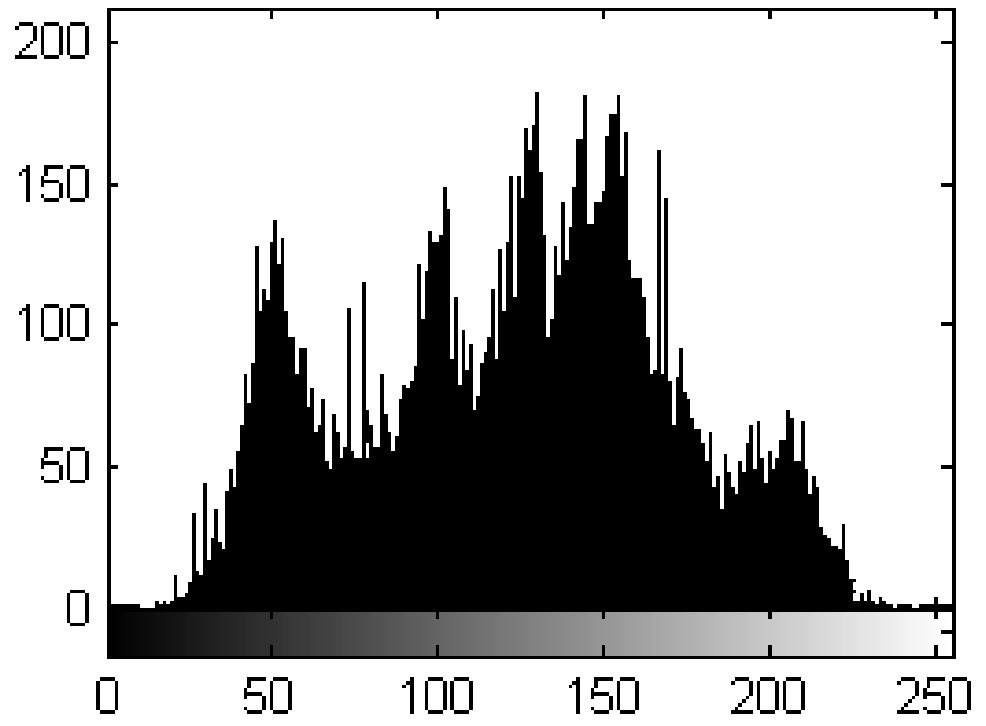

Figure 11. Histogram of sterilized Lena image

We see that our sterilization algorithm does not detectably distort the histogram of the input image.

## VI. Conclusion

In this paper, we have provided a novel concept of image sterilization. We have achieved on an average 76% to 91% success rate to sterilize the stego information of an image. We would like to emphasize that the goal of our technique is not hidden message recovery, rather we aim at annihilating stego information transmission without distorting the image visibly. Our approach is generic and applies to any LSB based steganography algorithm (Novel).

To our knowledge, this is the first work of its kind and so there does not exist any benchmark for comparing our technique with. We believe that our work would initiate interests in the community to pursue further research on this topic.